\documentclass[twocolumn, amsmath,amssymb,aps, pra,
showkeys,showpacs, floatfix, superscriptaddress] {revtex4-2}
\usepackage{times}
\usepackage{graphicx}
\usepackage[usenames,dvipsnames]{color}
\usepackage{bm}
\usepackage{amsmath}
\usepackage{amssymb}

\usepackage[colorlinks=true,breaklinks=true,allcolors=blue]{hyperref}

\def\re{{\rm e}}
\def\ri{{\rm i}}

\begin{document}

\title{Variational Quantum Simulation of Anyonic Chains} 
\author{Ananda Roy}
\email{ananda.roy@physics.rutgers.edu}
\affiliation{Department of Physics and Astronomy, Rutgers University, Piscataway, NJ 08854-8019 USA}

\begin{abstract}
Anyonic chains provide lattice realizations of a rich set of quantum field theories in two space-time dimensions. The latter play a central role in the investigation of generalized symmetries, renormalization group flows and numerous exotic phases of strongly-correlated systems. Here, a variational quantum simulation scheme is presented for the analysis of those anyonic chains which can be mapped to the restricted solid-on-solid~(RSOS) models of Andrews, Baxter and Forrester. An~$L_R$ site RSOS model associated with a Dynkin diagram containing~$p$ nodes is realized with~$L_R\lceil\ln_2 p\rceil$ qubits, where~$\lceil x\rceil$ is the smallest integer~$\geq x$. The scheme is benchmarked by realizing the ground states of RSOS Hamiltonians  in the~$A_p$ family for~$4\leq p\leq8$ using a variational quantum-classical algorithm. The latter is based on the Euler-Cartan circuit ansatz. Topological symmetry operators are analyzed for the RSOS models at the quantum-critical points. Measurement of observables acting on~$\lceil\ln_2 p\rceil$ qubits is shown to capture the anyonic nature of the Hilbert space. The described quantum simulation scheme provides a systematic approach to give rise to a large family of quantum field theories which have largely eluded physical realizations. 
\end{abstract}

\maketitle 

\section{Introduction}
One dimensional quantum spin chains, described by Hamiltonians acting on states residing in a Hilbert space with a tensor product structure, provide a fertile playground for the investigation of strongly-interacting quantum field theories~(QFTs). These QFTs are useful for modeling numerous physical systems ranging from magnetic materials~\cite{Heisenberg1928} and transitional metal oxides~\cite{Hase1993} to black holes~\cite{Ikhlef2012, Bazhanov:2020dlm}. These spin chains, being amenable to state-of-the-art numerical and sometimes analytical techniques, enable quantitative computation of several non-perturbative properties of the corresponding QFTs. 

Anyonic chains are nontrivial generalizations of the aforementioned spin chains, where the interacting spins are replaced by anyonic excitations~\cite{Feiguin:2006ydp, Gils2013}. The corresponding Hilbert space has no longer the tensor product structure characteristic of ordinary spin chains. Instead, the allowed states in this Hilbert space are determined by a set of fusion rules associated with a fusion category~(see, for example, Ref.~\cite{Buican:2017rxc}) with every allowed state corresponding to an admissible fusion diagram. The same fusion rules determine the anyonic chain Hamiltonian, which is built out of projectors favoring a set of fusion channels. The Hamiltonians commute with a set of operators which arise due to the existence of topological symmetries~\cite{Feiguin:2006ydp} -- a distinguishing feature of these anyonic chain models. In fact, these topological symmetry operators play a central role in the investigation of generalized symmetries in two-dimensional conformal field theories~(CFTs)~\cite{Aasen2016, Belletete2018, Aasen:2020jwb, Belletete2023, Sinha:2023hum}. 
\begin{figure}
\centering
\includegraphics[width=0.49\textwidth]{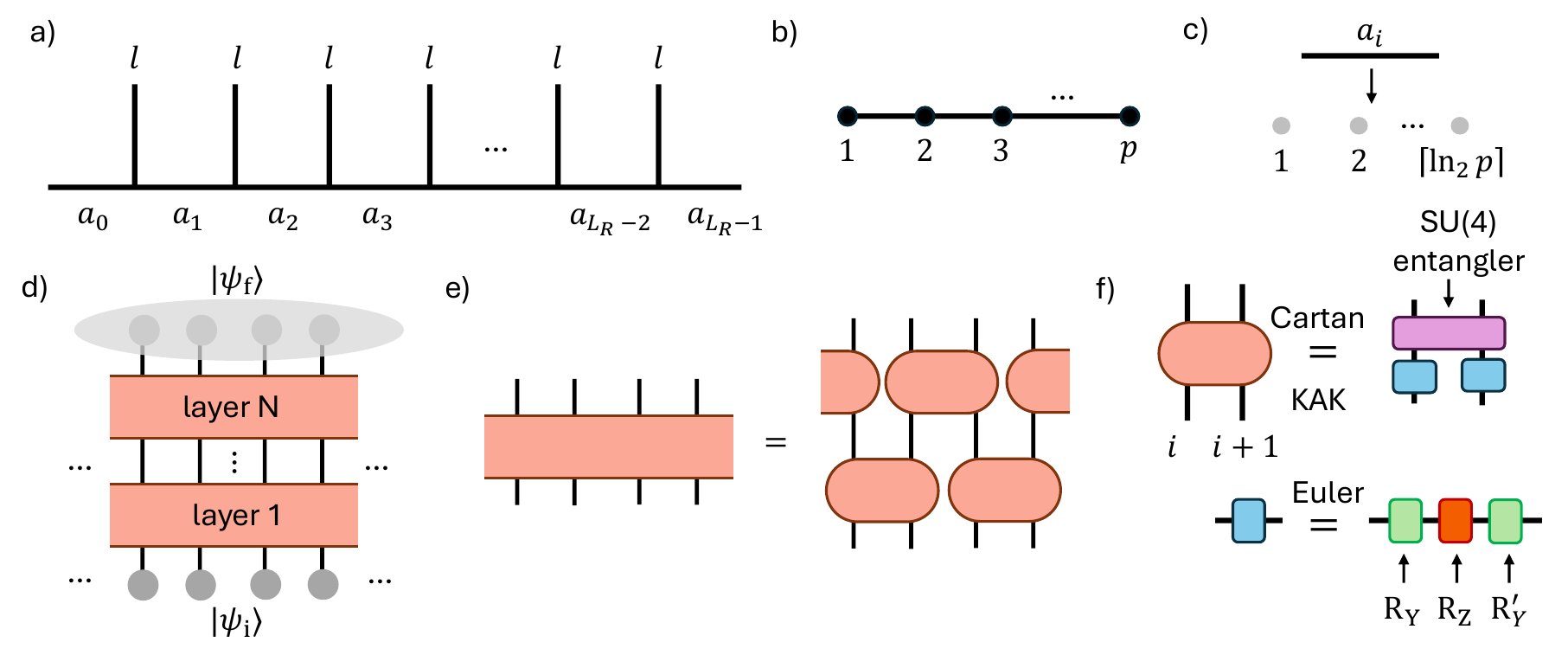}
\caption{\label{fig:schematic} a) Schematic of an anyonic chain, chosen to be translation invariant~(indicated by the same state on the vertical links). For the RSOS models, the states~$a_i$ can be one of the~$p$ states of the associated Dynkin diagram. b) Dynkin diagram~$A_p$,~$p=3,4,
\ldots$. The allowed states of the anyonic Hilbert space are such that~$a_i$ occurs in the fusion of~$a_{i-1}$ and~$l$. In the RSOS case, this amounts to saying that the element linking~$a_{i-1}$ and~$a_i$ in the adjacency matrix of~$A_p$ is nonzero. While the results are presented only for the A-type models, generalizations to the other types are straightforward. c) Encoding the $p$ states of the anyonic chain in~$\lceil \ln_2p\rceil$ qubits. d) Transformation of a product state~$|\psi_{\rm i}\rangle$ into the final state~$|\psi_{\rm f}\rangle$ after application of~$N$ layers of unitary operators. e) The parameterized circuit ansatz in terms of nearest-neighbor, potentially-distinct, compressed Euler-Cartan unitary operators. f) Decomposition of the latter in terms of single qubit rotations and an SU(4) entangler. While the former are represented by three Euler angles about the Y and Z axis, the latter is parametrized by~${\rm exp}\left(\ri\alpha_1 X_iX_{i+1} + \ri\alpha_2 Y_iY_{i+1} + \ri\alpha_3 Z_iZ_{i+1}\right)$.}
\end{figure}

In spite of their intriguing characteristics, physical realizations of anyonic chains are hard to come by. The lack of a tensor product structure of the anyonic Hilbert space, while responsible for their novel properties, also poses a significant challenge towards their physical realization. Simplifications arise in the cases of the Ising and three-state Potts models, which happen to have representations in terms of spin chains with tensor product Hilbert spaces. But, this is not true in general. Here, a systematic scheme is presented for mapping arbitrary anyonic chain Hamiltonians into those governing ordinary conventional SU(2) spins or qubits. The scheme is demonstrated for the restricted solid-on-solid (RSOS) models of Andrews, Baxter and Forrester~\cite{Andrews:1984af}. A given $L_R$-site RSOS model associated with a Dynkin diagram with $p$ nodes~\cite{Pasquier_1987, PASQUIER1987162, Saleur:1990uz} is embedded within a model of~$n_pL_R$ qubits, where~$n_p = \lceil\ln_2 p\rceil$ and~$\lceil x\rceil$ is the smallest integer~$\geq x$. The corresponding RSOS Hamiltonians, constructed with operators satisfying the Temperley-Lieb~(TL) algebra~\cite{Temperley1971}, act on~$3n_p$ neighboring qubits. The topological symmetry and braid operators, built with the same TL operators, are subsequently inferred in the qubit formulation. Clearly, the said embedding introduces additional `unphysical' states compared to those allowed by the RSOS models. However, the Hamiltonian, written in terms of projectors (see below), contains only those states in its eigenspectrum that satisfy the constraints arising from the adjacency matrix of the corresponding Dynkin diagram. Importantly, the proposed mapping allows the measurement of~$n_p$-qubit observables as a method for detection of the constraints that the physical states satisfy. 

The aforementioned mapping of RSOS models into those governing arrays of qubits enables their realization and investigation with digital quantum simulators~\cite{Feynman:85}. Given that canonical quantum algorithms like phase estimation~\cite{Shor_1994, Kitaev:1995qy} have limited near-term applicability, here, a variational quantum-classical algorithm~\cite{Peruzzo2011, Tilly2022} is proposed for the investigation of the qubit incarnation of the anyonic model. In this approach, starting from a trivial product state, a unitary operator defined by a parametrized circuit is determined to realize low-lying eigenstates of these RSOS models at their respective quantum critical points. The RSOS Hamiltonians, involving terms acting on~$3n_p$ qubits,~$n_p = 2, 3, \ldots$, make it difficult to apply some of the well-known operator ans\"{a}tze~\cite{Farhi2014, Grimsley2019} for the said unitary operator. Here, the unitary operator is parametrized using the compressed Euler-Cartan circuit ansatz for nearest-neighbor qubits~\cite{Roy2024ec}. The circuit parameters are determined using classical matrix-product state simulations of the circuit evolution and the ADAM optimization method~\cite{Kingma2017}. The latter is  implemented using PyTorch's reverse-mode automatic differentiation framework~\cite{Paszke2019}. Once the desired anyonic state is realized, the expectation value of the Hamiltonian, the topological symmetry operator and multi-qubit parity measurements that reveal the anyonic nature of the state space are computed. These are found to be in good agreement with the expected  results obtained using exact diagonalization and density matrix renormalization group~(DMRG) techniques. Note that the qubit formulation of the anyonic chain models also allows straightforward classical computation of their properties using the DMRG technique without explicitly conserving the anyonic charges~\cite{Sierra:1996qz, Pfeifer2015}. 

The paper is organized as follows. Sec.~\ref{sec:model} describes the reformulation of the RSOS models in terms of qubit operators. This is followed by description of the parametrized circuit ansatz in Sec.~\ref{sec:vqc}. Subsequently, results for the preparation of the ground state of RSOS chains with different boundary conditions are presented in Sec.~\ref{sec: results}. 
A concluding summary and outlook are provided in Sec.~\ref{sec:concl}. Appendix~\ref{sec:app_tci} provides an alternate representation of either the even or the odd sector of the tricritical Ising model using three-qubit interactions. 

\section{Qubit Hamiltonians for RSOS models}
\label{sec:model}
Some basic facts of RSOS models are collected below (for more details, see, for example, Ref.~\cite{Saleur:1990uz}). The allowed states in the Hilbert space for~$L_R$ sites are of the form~$|a_0, a_1, \ldots, a_{L_R - 2}, a_{L_R-1}\rangle$. Here, the values of any two neighbors~$a_i, a_{i+1}$ are such that there exists a link connecting the two corresponding nodes in the Dynkin diagram. 
The RSOS Hamiltonians are written in terms of the TL generators~$e_j$,~$j= 0,\ldots, L_R - 1$. The latter acts on three neighboring sites,~$a_{j-1}, a_j, a_{j+1}$ with matrix elements:
\begin{align}
\label{eq:e_j}
\langle \{a'_i\}|e_j|\{a_i\}\rangle &= \prod_{k\neq j}\delta_{a_k, a'_k}\frac{[\phi(a_j)\phi(a'_j)]^{1/2}}{\phi(a_{j-1})}\delta_{a_{j-1}, a_{j+1}},\\\label{eq:phi} \phi(a) &= \sqrt{\frac{2\gamma}{\pi}}\sin(a\gamma),~a = 1, \ldots, p,
\end{align}
where~$\gamma = \pi/(p + 1)$. Note that~$\phi(a)$ is the eigenvector of the adjacency matrix of the Dynkin diagram with the largest eigenvalue. The TL generators satisfy
\begin{equation}
\label{eq:TL_alg}
e_j^2 = (q + 1/q)e_j,\ e_je_{j\pm 1}e_j = e_j,
\end{equation}
with~$[e_j, e_k] = 0$ whenever $|j-k|>1$. Finally,~$q = \re^{\ri\gamma}$. Consider the Hamiltonian: 
\begin{equation}
\label{eq:H_TL}
H = -\frac{\gamma}{\pi\sin\gamma}\sum_{j = j_{\rm min}}^{j_{\rm max}}e_j,
\end{equation}
where~$j_{\rm min} = 0(1)$,~$j_{\rm max} = L_R - 1 (L_R - 2)$ for open~(periodic) boundary conditions. For a given~$p$, the low-energy properties of this Hamiltonian is described by the minimal CFT model~${\cal M}(p + 1, p)$  with~$p = 3, 4, \ldots$. 

Next, the mapping of the Hamiltonian~$H$ into one governing a register of qubits is described. The~$p$-states of the RSOS model are encoded into~$n_p = \lceil \ln_2p\rceil$ qubits such that the odd~(even) states contain an odd~(even) number of down spins. Equivalently, they can be distinguished by the operator~$\prod_{k = 1}^{n_p}Z_k$ acting on the~$n_p$ qubits. For example,
\begin{align}
p = 3:\nonumber\\\label{eq:p_3}|1\rangle &= |\downarrow\uparrow\rangle, |2\rangle = |\downarrow\downarrow\rangle, |3\rangle = |\uparrow\downarrow\rangle,\\
p = 4:\nonumber\\\label{eq:p_4}|1\rangle &= |\downarrow\uparrow\rangle, |2\rangle = |\downarrow\downarrow\rangle, |3\rangle = |\uparrow\downarrow\rangle, |4\rangle = |\uparrow\uparrow\rangle,\\
p = 5:\nonumber\\\label{eq:p_5}|1\rangle &= |\downarrow\downarrow\downarrow\rangle, |2\rangle = |\downarrow\downarrow\uparrow\rangle, |3\rangle = |\downarrow\uparrow\uparrow\rangle,\nonumber\\|4\rangle &= |\downarrow\uparrow\downarrow\rangle, |5\rangle = |\uparrow\downarrow\uparrow\rangle,
\end{align}
and so on. Thus, an~$L_R$ size RSOS model requires~$L = n_pL_R$ number of qubits. The allowed states in the RSOS Hilbert space contain RSOS sites with alternating odd and even parity. When translated to the qubit language, this manifests in an oscillating behavior of the aforementioned~$n_p$-qubit parity operator (see also Fig.~\ref{fig_2}). The anyonic nature of the Hilbert space imposes additional restrictions on the set of allowed states in addition to what is captured by the~$n_p$-qubit parity measurement. This is because, if a given site is in state~$|a\rangle$, for the class of models analyzed here, the only allowed states are~$|a\pm1\rangle$. For example, if a given RSOS site is in the state~$|2\rangle$, its neighbor has to be in the state~$|1\rangle,|3\rangle$ and no other possibilities are allowed. However, if a site is in the state~$|1\rangle$, its neighbor can only be in the state~$|2\rangle$. This additional constraint can be verified by computing the probability of the different RSOS sites being in the state~$|a\rangle$,~$a = 1, \ldots, p$.  This again amounts to measurement of observables acting on~$n_p$ neighboring qubits, see Eq.~\eqref{eq:P_a} below and Fig.~\ref{fig_2}. 

Note that in the case of the Ising model, a more efficient encoding exists using the conventional spin chain representation. This is because, the even state~$|2\rangle$ is always fixed to the same value in the RSOS Hilbert space, allowing for the states~$|1\rangle,|3\rangle$ to be encoded into the two states of a single qubit. A similar encoding is also provided for the tricritical Ising model in Appendix~\ref{sec:app_tci} involving three-spin interactions. 

With this mapping, the TL generators can be written as: 
\begin{align}
\label{eq:TL_op}
e_j = \sum_{a = 1}^p P_{j-1}^{(a)}\tilde{e}^{(a)}_j P_{j +1}^{(a)},
\end{align}
where
\begin{align}
\label{eq:P_a}
P_j^{(a)} &= \frac{1}{\sqrt{\phi(a)}}|a\rangle\langle a|,\\
\tilde{e}_j^{(a)} &= \sum_{b_1, b_2\in {\cal N}_a}\sqrt{\phi(b_1)\phi(b_2)}|b_1\rangle\langle b_2|,
\end{align}
where~$\phi(a)$ is defined in Eq.~\eqref{eq:phi} and~${\cal N}_a$ denotes the set of nodes that are adjacent to~$a$ in the corresponding Dynkin diagram. Notice that each of the TL generators acts on~$3n_p$ qubits. The explicit forms of the TL generators are not provided, except for the case in Appendix~\ref{sec:app_tci}, since they are not particularly illuminating. That the said representation satisfies TL algebra~[Eq.~\eqref{eq:TL_alg}] was verified by explicitly multiplying the operators of Eq.~\eqref{eq:TL_op}. Given that the TL generators are identified in terms of qubit operators, all the Virasoro generators can then be obtained using the Koo-Saleur formula~\cite{Koo_1994}. The corresponding braid generators can also be written in terms of the qubit operators using~\cite{Saleur:1990uz}:
\begin{equation}
g_j = (-q)^{1/2}\left(1 - e_j/q\right),
\end{equation}
which satisfy
\begin{align}
g_jg_{j\pm1}g_j &= g_{j\pm1}g_jg_{j\pm1},\\ g_jg_{k} &= g_{k}g_j,\ |j-k|\geq2. 
\end{align}
Finally, in this formulation, the topological symmetry operator is given by~\cite{Sinha:2023hum}
\begin{equation}
\label{eq:Y_def}
Y = (-q)^{-1/2}g_0^{-1}\ldots g_{L_R - 2}^{-1}u^{-1} + {\rm h.c.},
\end{equation}
where~$u$ is the shift operator that performs translation of the RSOS chain by one site. Equivalently, it translates the lattice of qubits by~$n_p$ sites. Note that this definition of the topological symmetry operator is equivalent to the one defined using the corresponding fusion amplitudes in Ref.~\cite{Feiguin:2006ydp}. In fact, the operator~$Y$ corresponds to the topological defect line associated with the Kac label~$(1,2)$ of the corresponding minimal model~${\cal M}(p + 1, p)$~\cite{Petkova:2000ip}. As such, its expectation value on the ground state of a periodic RSOS chain is given by the ratio of the corresponding S-matrix elements:~$S_{12, 11}/S_{11, 11}$. For the model~${\cal M}(p + 1, p)$, this expectation value is given by~$2\cos[\pi/(p + 1)]$~\cite{DiFrancesco:1997nk}. This expectation value serves as an additional diagnostic for benchmarking the fidelity of the state obtained using the variational quantum circuits~(see also Fig.~\ref{fig_3}). 

\section{Variational Circuits for RSOS models}
\label{sec:vqc}
Next, a parametrized quantum circuit ansatz is described which, upon suitable optimization procedure, transforms an unentangled state to the ground state of the RSOS Hamiltonians. While the optimization procedure typically relies on variations of gradient descent methods~\cite{Fletcher2013, Kingma2017,Stokes2020},  there is no natural prescription for the choosing the quantum circuit ansatz. Typical approaches are based on the evolution by the Hamiltonian whose eigenstate is being sought alongside a heuristically-chosen mixer Hamiltonian~\cite{Farhi2014, Hadfield2019, Wang2020} or variations of the coupled cluster theory relevant mostly for chemistry problems~\cite{Peruzzo2011, Grimsley2019, Tang2021}. Neither of the aforementioned are suitable for the RSOS Hamiltonians which involve~$3n_p$-qubit interactions. The problem is made more difficult by the fact that the described models are tuned to their quantum-critical points with the corresponding ground states exhibiting similar average values for the energy and containing similar amount of entanglement. 
\begin{figure}
\centering
\includegraphics[width=0.49\textwidth]{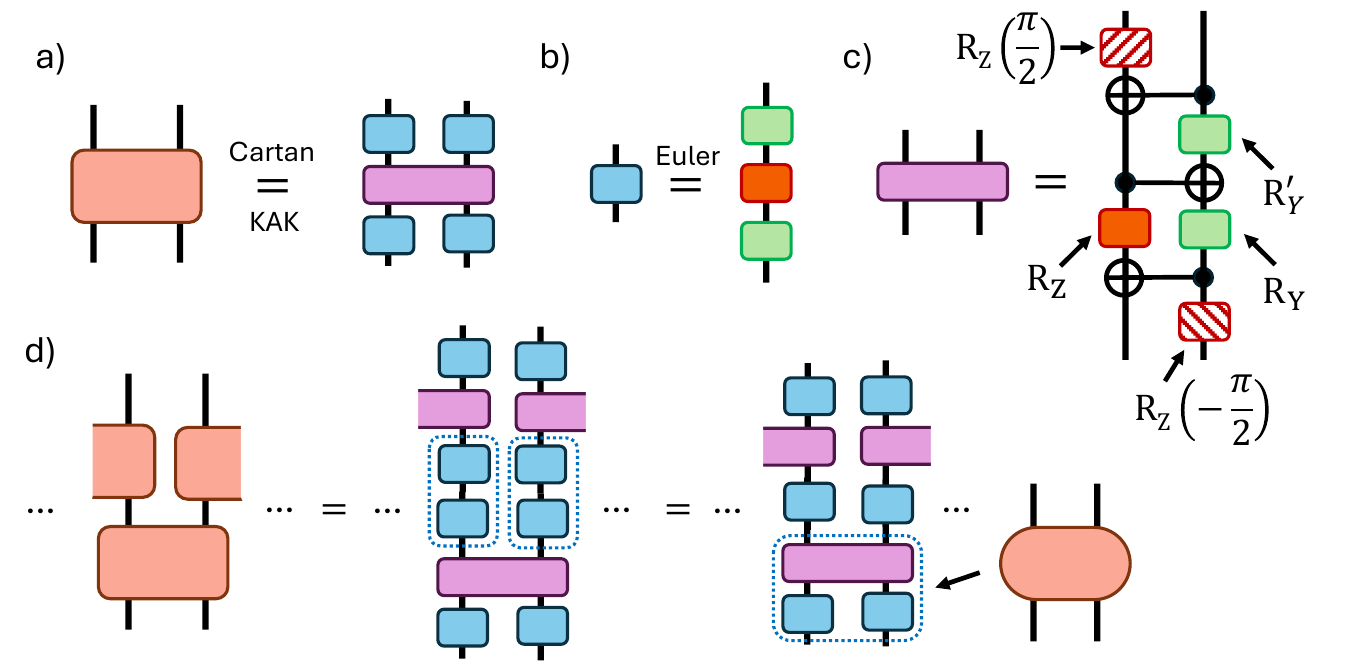}
\caption{\label{fig:KAK} a) Cartan's KAK decomposition of an SU(4) operator in terms of four single qubit rotations and an SU(4) entangler. b) Euler decomposition of a single qubit rotation into three rotations about two non-parallel axis (here chosen to be about the Y and Z axes). c) Decomposition of the SU(4) entangler into rotations about Y and Z axes and three CNOT gates. d) Steps for going from the KAK decomposed circuit of nearest-neighbor SU(4) unitary operators to the circuit ansatz of Fig.~\ref{fig:schematic}. }
\end{figure}

Here, the recently-proposed Euler-Cartan circuit ansatz for nearest-neighbor qubits~\cite{Roy2024ec} is used. The geometric locality is imposed to facilitate implementation in majority of quantum computing architectures which are realized using locally-interacting qubits. At the same time, this allows numerical benchmarking of the proposed scheme using the time-evolved block-decimation~(TEBD) algorithm~\cite{Vidal2004tebd} for matrix product states. Allowing longer range couplings would likely lead to faster buildup of entanglement and further improve the efficacy of the scheme. The Euler-Cartan ansatz is based on the KAK decomposition of a general SU(4) rotation in terms of four single-qubit rotations by Euler angles and an SU(4) entangler~[see Fig.~\ref{fig:KAK}a)]. The latter can be parameterized by three parameters~$\alpha_i$,~$i = 1, 2, 3$ as: ${\rm exp}\left(\ri\alpha_1 X_iX_{i+1} + \ri\alpha_2 Y_iY_{i+1} + \ri\alpha_3 Z_iZ_{i+1}\right)$. These three parameters, together with the twelve Euler angles [from the four single qubit rotations of Fig.~\ref{fig:KAK}a),b)], comprise the set of fifteen parameters of an SU(4) rotation~\cite{Vidal2004, Vatan2004}. For the parametrized circuit ansatz, two of the single-qubit rotations of an SU(4) operator can be combined with those of another in the next sublayer~[Fig.~\ref{fig:KAK}d)] leading to the circuit ansatz of Fig.~\ref{fig:schematic}e), f). For implementation on a quantum computer, it is useful to further decompose the SU(4) entangler into single qubit rotations and three CNOT gates as shown in Fig.~\ref{fig:KAK}c). This leads to nine parameters per two-qubit rotation~[see Fig.~\ref{fig:KAK}c),d)] that need to be determined during the optimization procedure. 
\begin{figure}
\centering
\includegraphics[width=0.49\textwidth]{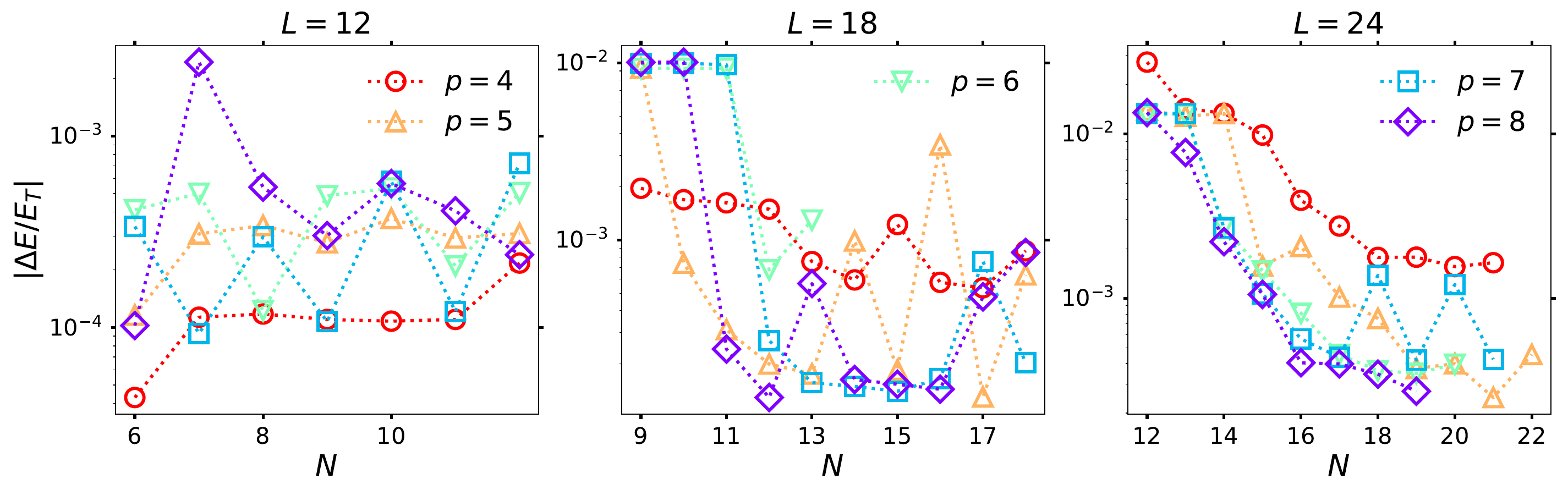}
\caption{\label{fig_1} Errors in the energy~$|\Delta E/E_T| = |E/E_T - 1|$ obtained by minimizing the expectation value of the Hamiltonian~[Eq.~\eqref{eq:H_TL}] as a function of the number of layers~($N$) for~$4\leq p\leq 8$. The target energy~$E_T$ is obtained from exact diagonalization of the RSOS model. The results are shown for 12~(left), 18~(center) and 24~(right) qubits with open boundary conditions. The number of layers was increased in steps of 1 starting with $N = L/2$. The initial guess for all the angles~[see Fig.~\ref{fig:schematic} f) and Fig.~\ref{fig:KAK} d)] was chosen to be~$\theta_0 = 1.0$ for~$N = L/2$. To improve stability of the optimization procedure, results obtained for the $(N - 1)^{\rm th}$ layer was used, together with the choice of~$\theta_0/10$ for the added layer. The number of layers was increased even after the target relative error of~$5\times10^{-3}$ was reached to test the stability of the simulation scheme. }
\end{figure}

\begin{figure*}
\centering
\includegraphics[width=\textwidth]{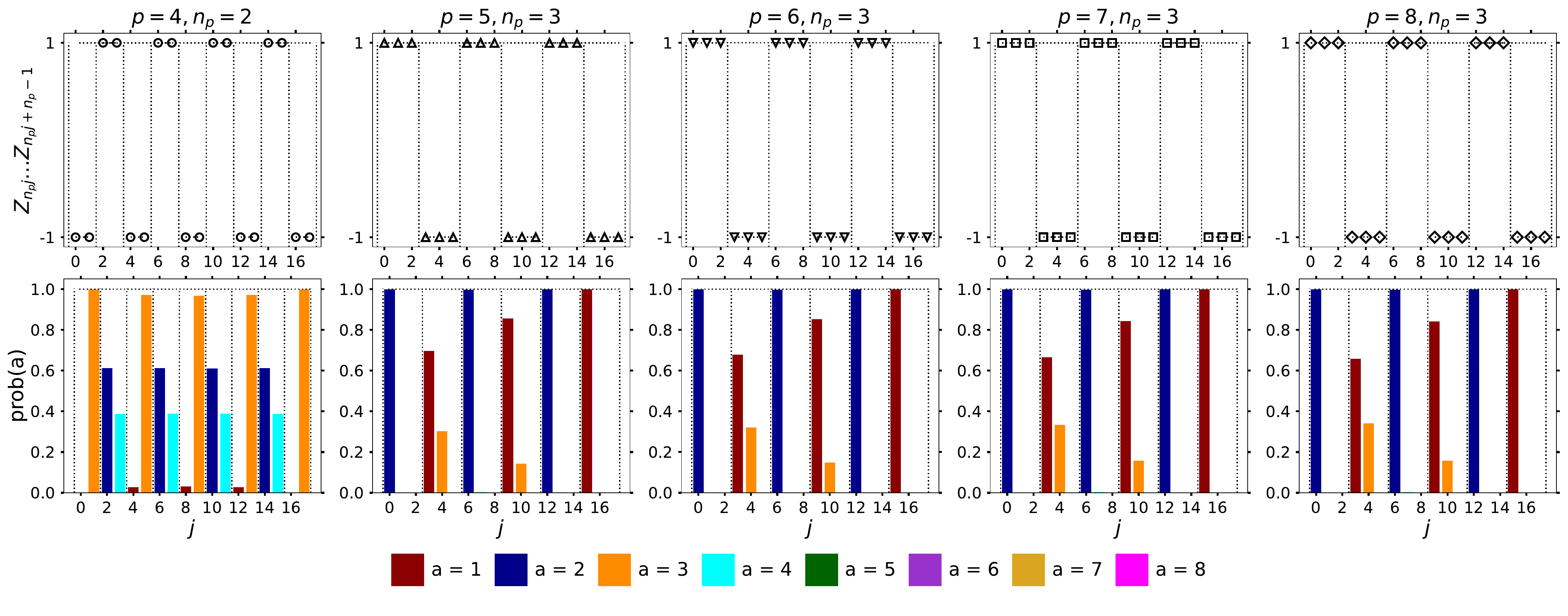}
\caption{\label{fig_2} Expectation values of observables computed from the states obtained in Fig.~\ref{fig_1}. Results are shown for~$L = 18$ with open boundary conditions. The dashed lines indicate the boundaries of the RSOS sites built out of~$n_p=2$ ($p = 4$) and 3 ($5\leq p\leq8$) qubits. (Upper panels) Expectation values of the~$n_p$-qubit parity operator~$\prod_{k = n_p j}^{n_p j + n_p - 1}Z_k$ along the chain. The alternating~$\pm 1$ values for the said parity operator is expected of any allowed state in anyonic RSOS Hilbert space.~(Lower panels) The probabilities of the RSOS site being in the state~$a$, denoted by~$p(a)$ are shown, where~$a = 1, 2, \ldots, p$. This is computed by evaluating the expectation value of the operator~$|a\rangle\langle a|$ for the~$j^{\rm th}$ RSOS site. In addition to each RSOS site being in the correct parity sector, each is only in the set of states allowed by the Dynkin diagram of Fig.~\ref{fig:schematic}b). The probabilities were found to be compatible with DMRG results. Note that for~$L = 18$ qubits, for~$p = 4~(p > 4)$, the RSOS model has odd~(even) number of sites. This leads to the boundary spins being fixed to the same~(different) values for~$p = 4~(p > 4)$. The values of the boundary spins were chosen by suitably choosing the initial state during the optimization procedure.}
\end{figure*}

Below, results are presented to benchmark the proposed scheme by realizing ground states of different RSOS models with different boundary conditions. The optimization is performed by implementing the~(unitary) circuit evolution using the TEBD algorithm for matrix product states. For the ease of implementation, the SU(4) entangler between the first and the last sites that arises during the second sublayer of a given layer~[see Fig.~\ref{fig:schematic}e)] is dropped. Thus, for $N$ layers and $L$ qubits, the total number of parameters to be determined by the optimization procedure is~$6LN + 3(L - 1)N$. Here, the~$6LN$ parameters arise from single qubit rotations and the rest from the SU(4) entanglers. The associated optimization is performed using the ADAM optimizer~\cite{Kingma2017}, implemented using PyTorch's reverse-mode automatic differentiation framework~\cite{Paszke2019}. The latter requires some modifications when used during the singular value decomposition~(SVD) of complex matrices necessary for unitary evolution of quantum states. A Lorentzian broadening is used to avoid divergences associated with near-degenerate singular values~\cite{Liao2019}. The associated broadening parameter was set to~$10^{-12}$. Furthermore, the presence of complex matrices requires the inclusion of an extra term~\cite{wan2019automaticdifferentiationcomplexvalued} in the computation of the gradient used in backpropagation during the SVD truncation step. This is in addition to the usual ones discussed in Refs.~\cite{Townsend2016, seeger2019autodifferentiatinglinearalgebra}. Finally, for time-evolution, the singular values below~$10^{-10}$ were dropped. The bond-dimension for the TEBD algorithm was set to sufficiently high values so that the associated truncations did not alter the results obtained up to six decimal places. 

\section{Results for ground states of RSOS models}
\label{sec: results}
First, results are presented for RSOS models with open boundary conditions. In this case, the boundary RSOS spins are fixed to a certain value and the corresponding models are described the minimal models~${\cal M}(p + 1, p)$ with appropriate boundary conditions~\cite{SALEUR1989591}. The Hamiltonian in Eq.~\eqref{eq:H_TL} in fact realizes the boundary condition~$(1, s)$~$s = 1, \ldots, p$, although this is not important for this work. 

Fig.~\ref{fig_1} shows the relative errors in the energies obtained by minimizing the expectation value of the Hamiltonian of Eq.~\eqref{eq:H_TL} for systems with~$L = 12, 18$ and 24 qubits. The target energy,~$E_T$, is obtained from exact diagonalization of the said Hamiltonian. The initial state for the quantum circuit optimization was chosen to be a RSOS state~$|2, 1, 2, 1, \ldots\rangle$. The corresponding states of the qubit register can be obtained using the mapping of Eqs.~(\ref{eq:p_3}-\ref{eq:p_5}). The optimization scheme was found to be not sensitive to the choice of the initial state. The system-size of the RSOS model can be inferred using the fact that for~$p = 4~(5\leq p\leq8)$,~$n_p = 2(3)$. The number of layers (equivalently, the circuit depth) was systematically increased from~$N = L/2$ in steps of 1. The initial guess for all of the undetermined angles was chosen to be~$\theta_0 = 1.0$. To enhance the stability of the optimization process, the results obtained from optimization of~$N - 1$ layers was used in the~$N^{\rm th}$ layer, augmented by an initial guess of~$\theta_0/10$ for the angles of the new layer. The average value of the energy was computed by computing relevant correlation functions~[Eq.~\eqref{eq:TL_op}] of ~$3n_p$ qubits. In all the cases analyzed, the errors in the energies were below~$5\times10^{-3}$ within a circuit depth of~$N \simeq L$ layers. This is compatible with earlier findings~\cite{Roy2023efficient, Rogerson2024} where the circuit-depth was found to scale proportional to the system-size for realization of gapless ground states. 

In addition to the average energy, several observables that capture the anyonic nature of the Hilbert space were computed with the states obtained after the optimization process. The upper panels of Fig.~\ref{fig_2} show the joint-parity measurements for~$n_p$ neighboring qubits for~$L = 18$ and~$4\leq p\leq8$. Since an RSOS site with odd~(even) states can have even~(odd) states in the neighboring sites~(Sec.~\ref{sec:model}), the~$n_p$-qubit parity is expected to oscillate between~$+1$ and~$-1$ values. This is verified in Fig.~\ref{fig_2}~(upper panels). The lower panels show the probability~$p(a)$ of the RSOS spin being in the state~$a = 1, \ldots, p$. The latter captures the additional constraints of the anyonic Hilbert space of the RSOS models. As verified in the different panels,~$p(a)$ is nonzero for only those states that are allowed by the corresponding Dynkin diagram~(and not all of those who have the correct~$n_p$-qubit parity). For instance, for~$p = 4$, an~$L = 18$ chain of qubits realizes an open 9-site RSOS chain with fixed spins at the ends. Fig.~\ref{fig_2}~(bottom left) shows the results where~$p(a=3)$ at the ends with their neighbors having~$p(a = 2)$ and~$p(a = 4)$ being nonzero. In contrast, for~$5\leq p\leq 8$, the qubit-chain realizes a 6-site RSOS model with the left and right boundary spins fixed to the states~$|2\rangle$ and~$|1\rangle$ respectively. Then, the second from left RSOS site has~$p(a = 1)$ and~$p(a = 3)$ nonzero. However, the second from right RSOS site has~{\it only}~$p(a = 2)$ nonzero~(see Fig.~\ref{fig_2}, the last four bottom panels). 

\begin{figure}
\centering
\includegraphics[width=0.49\textwidth]{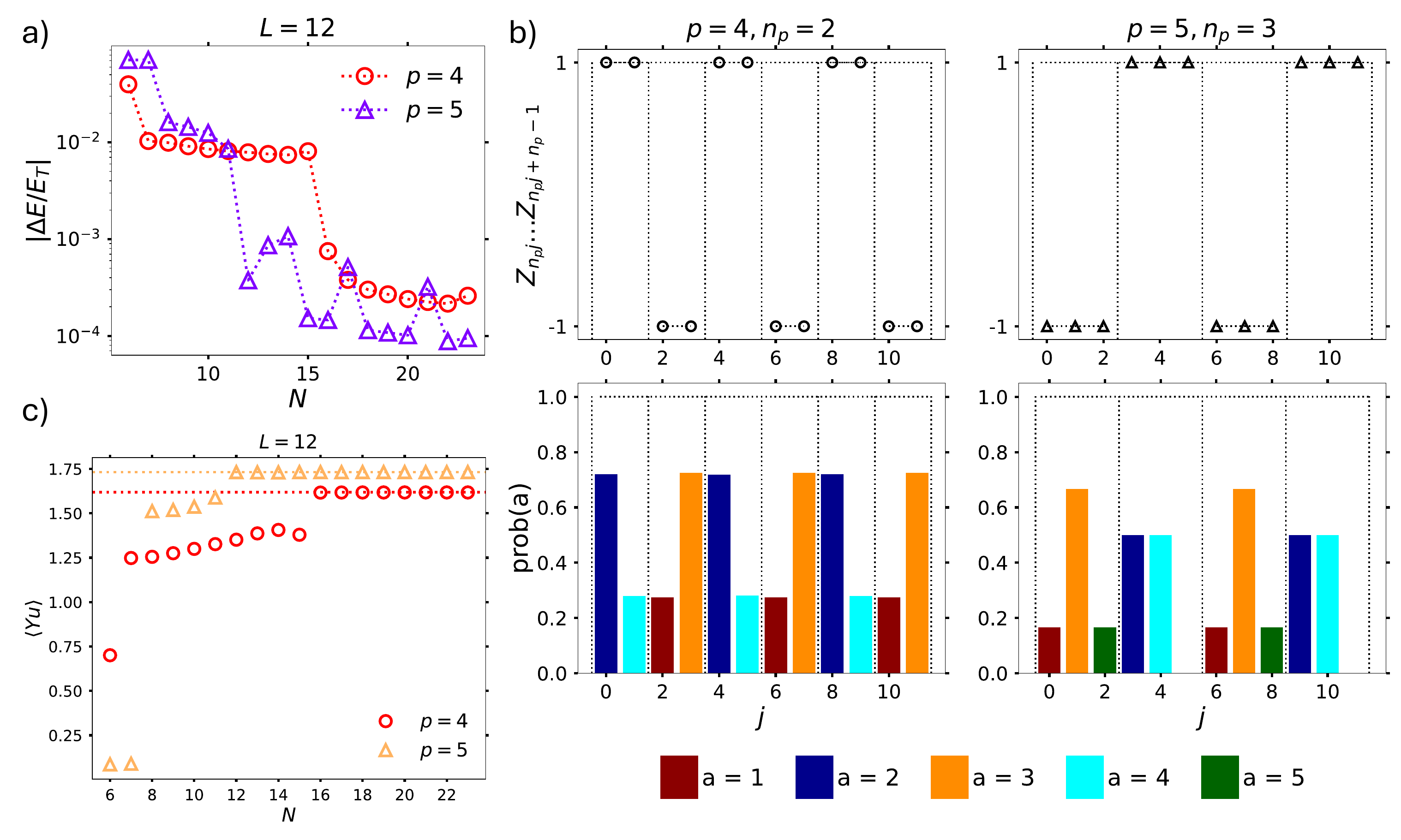}
\caption{\label{fig_3} a) Errors in the energy~$|\Delta E/E_T| = |E/E_T - 1|$ obtained by minimizing the expectation value of the Hamiltonian~[Eq.~\eqref{eq:H_TL}] as a function of the number of layers~($N$) for~$p = 4, 5$. The target energy~$E_T$ is obtained from exact diagonalization of the RSOS model. The results are shown for 12 qubits with periodic boundary conditions. The initial conditions and the optimization strategies are identical to Fig.~\ref{fig_1}. b) (Upper panels) Expectation values of the~$n_p$-qubit parity operator~$\prod_{k = n_p j}^{n_p j + n_p - 1}Z_k$ along the chain.~(Lower panels) The probabilities~$p(a)$ of the RSOS site being in the state~$a$,~$a = 1, 2, \ldots, p$. In contrast to the open boundary conditions~(Fig.~\ref{fig_1}, bottom panels), the state of the RSOS spin at the ends of the chain are no longer pinned a fixed value. Instead they are restricted to a given parity sector. c) Expectation values of~$Yu$ in the ground states of RSOS models with periodic boundary conditions. Here,~$Y$ is the topological symmetry operator~[Eq.~\eqref{eq:Y_def}] and~$u$ is the shift operator that performs translation by~$n_p$ qubits. Results are shown for~$L = 12$ and~$p=4,5$. The obtained values for~$p=4$ and $5$ are 1.6177 and 1.7319 respectively, which are close to the expected results given by~$2\cos[\pi/(p + 1)]$. }
\end{figure}

The above results can be contrasted for the RSOS chain with periodic boundary conditions. In this case, all the RSOS spins are allowed to fluctuate, albeit within a given parity sector~[see the discussion below Eq.~\eqref{eq:p_5}]. Note that this difference makes the optimization process more arduous. However, the same optimization strategy as in the case of open boundary conditions can still be used with the gradual increase of the number of layers from~$N = L/2$ in steps of 1 and the dropping of the boundary~SU(4) entanglers. 

Fig.~\ref{fig_3}a) shows the results for the relative error in average energy for~$L = 12$ for~$p = 4, 5$~(similar results were obtained for~$L = 18$ and higher values of~$p$, but are not shown for brevity). As shown, the errors in the obtained energies are within the threshold of~$5\times10^{-3}$. Fig.~\ref{fig_3}b) shows the results for the~$n_p$-qubit parity measurements~(upper panels) and the occupation probabilities,~$p(a)$,~$a = 1, \ldots, p$~(bottom panels). The results for the~$n_p$-qubit parity measurements show the same oscillatory behavior as obtained for open boundary conditions due to the even-odd occupancy of the RSOS sites. However, the occupation probabilities are indeed different. This is because the boundary spins are no longer pinned to a given value. As a result more than one value of~$a$ is allowed for each of the RSOS sites~[compare bottom panels of Figs~(\ref{fig_2}, \ref{fig_3})]. Finally, Fig.~\ref{fig_3}c) shows the results for the computation of the expectation value of the topological symmetry operator in the ground state of the periodic RSOS chain. As the number of layers are increased, the said value saturates at the desired value of~$2\cos[\pi/(p + 1)]$. Note that due to the even-odd parity of the RSOS Hilbert space, what is actually measured is the expectation value of~$Yu$, which has the same expectation value as the~$Y$ operator. 

\section{Summary and Outlook}
\label{sec:concl}
To summarize, this work provides a systematic way to realize anyonic chains with variational quantum algorithms in near-term quantum simulators. This is demonstrated for those chains which can be mapped to RSOS models with the A-type Dynkin diagram. When mapped to qubits, the resulting models involve~$3\lceil\ln_2p\rceil$-qubit interactions, with~$p = 3, 4, \ldots$. In contrast to the Ising and Potts chains which have natural representations in terms of Hilbert spaces with a tensor product structure, the physical states of these anyonic models satisfy specific constraints arising from the corresponding fusion rules. These constraints are imposed in the variational algorithm through the Hamiltonian cost-function using suitable projectors. Indeed, the mapping to the qubit registers introduces unphysical states that do not satisfy the constraints from the anyonic fusion rules. However, the states obtained using the quantum circuit optimization with the suitable cost function do obey the relevant constraints. The circuit ansatz is chosen to be the Euler-Cartan ansatz for nearest-neighbor interactions and the corresponding parameters are determined using the ADAM optimization method. Results are presented the RSOS models at their quantum critical points for~$4\leq p\leq8$. These critical ground states have larger entanglement than their gapped counterparts. The obtained circuit-depths are found to grow proportional to the system-size, which is compatible with earlier findings. Measurement of~$\lceil\ln_2p\rceil$-qubit observables are shown to capture the signatures of the anyonic nature of the Hilbert space and the corresponding topological symmetry operators. 

The proposed approach to realize anyonic chains enables quantum simulation of a large number of models which are difficult to give rise to using other physical systems. Some potential future research directions are outlined next. Corresponding to the topological symmetry operator, an impurity Hamiltonian can be constructed in the so-called `direct channel'. The resulting model is described by a CFT coupled to an impurity and provides  generalizations of multi-channel Kondo model~\cite{Ludwig1994, Affleck1995conformal, Fendley1999, Bachas:2004sy}. The described scheme can be used to investigate various questions in impurity scattering including the RG flow between topological defect fixed points in a given CFT~\cite{Kormos:2009sk, Tavares:2024vtu}. Given that several of the bulk primary fields of the CFTs discussed in this work have already been identified for the RSOS models~\cite{Pasquier_1987, PASQUIER1987162, Koo_1994}, the proposed scheme would enable quantum simulation of bulk  RG flows between successive minimal models~\cite{Zamolodchikov1986, Zamolodchikov1991a, Fendley1993}. Further generalizations of the discussed models lead to parafermionic CFTs~\cite{Fateev:1985mm} and their perturbations~\cite{FATEEV199191}. Finally, similar mappings can be used for not only generic~$s>1/2$ spin-chains, but also non-compact ones~\cite{Bytsko2006}. Given the pace of progress in quantum hardware development, there are reasons to be optimistic about large-scale investigation of non-perturbative phenomena in low-dimensional QFTs in the near future. 

\section*{Acknowledgments}
Discussions with David Rogerson and Hubert Saleur are gratefully acknowledged. 
\appendix
\section{Alternate encoding of the tricritical Ising model}
\label{sec:app_tci}
In this section, an alternate encoding of the RSOS tricritical Ising model~(corresponding to the Dynkin diagram~$A_4$) is described. Each site, in principle, can be occupied by the four states~$a = 1, 2, 3, 4$. However, due to the the constraints in the Hilbert space arising from the connectivity of~$A_4$, each site of the said model can be either in the states~$|1\rangle, |3\rangle$ or the states~$|2\rangle, |4\rangle$. Consider for definiteness the case when the odd~(even) sites are in the odd~(even) states. Then, an effective spin degree of freedom can be defined on the odd and even sites as: 
\begin{align}
|1\rangle_{\rm odd} &= |\Downarrow\rangle,~|3\rangle_{\rm odd} = |\Uparrow\rangle, \\ |2\rangle_{\rm even} &= |\Uparrow\rangle,~|4\rangle_{\rm even} = |\Downarrow\rangle. 
\end{align}
Notice that with this encoding, the Hilbert space constraint arising from the Dynkin diagram~$A_4$ reduces to the constraint that the spin on the right of a down spin is always in the up state. No such constraint arises for the spin on the right of an up spin. The said constraint is `unidirectional': no constraint exists for the spin on the left of an up spin. Equivalently, each allowed state,~$|\psi\rangle$, of the RSOS Hilbert space satisfy: 
\begin{equation}
P_j^-P_{j+1}^-|\psi\rangle = 0,
\end{equation}
where $P_j^\pm = (1\pm Z_j)/2$. With this encoding, it is straightforward to see that 
\begin{align}
P_j^{(1)} &= P_j^{(4)} = \frac{1}{\sqrt{\lambda}}P_j^-,\\ P_j^{(2)} &= P_j^{(3)} = \frac{1}{\sqrt{\mu}}P_j^+, 
\end{align}
where~$\lambda = \phi(1)$, $\mu = \phi(2)$~[see Eq.~\eqref{eq:phi}]. Similarly,
\begin{align}
\tilde{e}_j^{(1)} &= \tilde{e}_j^{(4)} = \mu P_j^+, \\
\tilde{e}_j^{(2)} &= \tilde{e}_j^{(3)} = \lambda P_j^- + \mu P_j^+ + \sqrt{\lambda\mu}X_j.
\end{align}
Thus, the RSOS Hamiltonian is given by Eq.~\eqref{eq:H_TL} with the TL generators:
\begin{align}
e_j &= \frac{\mu}{\lambda}P_{j-1}^-P_{j}^+P_{j+1}^-\nonumber\\&\quad + \frac{1}{\mu}P_{j-1}^+\left(\mu P_j^+ + \lambda P_j^- + \sqrt{\lambda\mu}X_j\right)P_{j+1}^+.
\end{align}
Note that the resulting Hamiltonian acting on~$L$ RSOS sites requires~$2L$ qubits instead of~$4L$ as proposed in Sec.~\ref{sec:model}. Similar encodings can be designed for all models associated with the diagrams~$A_p$,~$p>4$ as well, but not always possible for D and E class of models. 

\bibliography{/Users/ananda/Dropbox/Bibliography/library_1}

\end{document}